\documentclass[prb,twocolumn,floatfix,showpacs,superscriptaddress]{revtex4-1}
\usepackage{graphicx,multirow}

\begin{document}

\title{A van der Waals DFT study of chain length dependence 
of  \\ alkanethiol adsorption on Au(111): \\ Physisorption vs.chemisorption}

\author{Ersen Mete}\email{emete@balikesir.edu.tr}
\affiliation{Department of Physics, Bal{\i}kesir University, 
Bal{\i}kesir 10145, Turkey}
\author{Merve Yortanl{\i}}
\affiliation{Department of Physics, Bal{\i}kesir University, 
Bal{\i}kesir 10145, Turkey}
\author{Mehmet Fatih Dan{\i}\c{s}man}\email{danisman@metu.edu.tr}
\affiliation{Department of Chemistry, Middle East Technical 
University, Ankara 06800, Turkey}

\begin{abstract}
The energetics and structures of physisorbed and 
chemisorbed alkanethiols on Au(111) have been systematically investigated 
up to 10 carbon atoms using van der Waals (vdW) corrected density functional 
theory (DFT) calculations. The role of chain length, tilting angle and coverage 
on the adsorption characteristics have been examined to elucidate the 
energetics and plausible transformation mechanisms between lying down and 
standing up phases. Coverage and size dependent chain-chain electronic 
interactions counteract with the alkyl chain-gold surface interactions and the 
surface relaxation of the metal in the formation of standing up monolayer 
structures. For the striped phases of long chain alkanethiols, however, our 
calculations on decanethiol indicates alkyl chain-gold surface interactions to 
be strong enough to force the molecule to be perfectly parallel to the surface 
by lifting a gold atom up, in agreement with the proposed models for this film 
in the literature. 
\end{abstract}

\maketitle

\section{Introduction}

Thiol self-assembled monolayers (SAMs) on Au(111) surfaces continue to attract 
considerable interest due to their uses in many different applications ranging 
from organic electronics to biotechnology.\cite{matharu,claridge} These systems 
have been thoroughly investigated both experimentally and theoretically since 
the pioneering studies in 80's.\cite{schreiber1,love,porter,bain} 
Nevertheless, there are still issues, regarding their fundamental properties, 
which are not resolved completely.\cite{vericat1,maksymovych,woodruff} To 
tackle and clarify these issues is not only important from a fundamental 
surface science point of view but also necessary to improve and 
control/manipulate the properties of the applications/devices that rely on such 
SAMs.

The crystal structure of alkanethiol [CH$_3$(CH$_2$)$_{n-1}$SH, will be 
referred to as C$n$] SAMs are well established experimentally and it is known 
that depending on the film density these SAMs have several phases. At low 
density, the so called striped phases, made up off lying down molecules, form. 
These have the general structure of ($p\times\sqrt{3}$) rectangular unit cells 
where the periodicity, $p$, which depends on the film density and the length of 
the molecule, gives the stripe separation along the gold nearest neighbor 
direction. Decanethiol is the most studied thiol SAM system and for its lowest 
density films it has a (11$\times\sqrt{3}$) unit cell [some groups report a 
(11.5$\times\sqrt{3}$) unit cell instead] which is also referred to as $\beta$ 
phase. Upon increasing coverage a new striped phase forms with a 
(7.5$\times\sqrt{3}$) structure which is also referred to as $\delta$ 
phase.\cite{poirier,qian,camillone,albayrak,toerker,darling} 
Hexanethiol, another well studied molecule, on the other hand has a (7.5 
$\times\sqrt{3}$) striped phase.\cite{camillone,kondoh1,dubois,shimada} Similar 
striped phases were observed for other chain lengths with p values in 
accordance with the length of the thiol molecule.\cite{schreiber1,love} For the 
highest density films ($\theta$=1/3) molecules stand up and form the well-known 
($\sqrt{3}\times\!\sqrt{3}$) R30$^\circ$ unit cell structure with a 
$c$(4$\times$2) superlattice, regardless of the chain 
length.\cite{schreiber1,dubois,cossaro,wang,kondoh2} However, for 
shorter chain lengths (C1-C3), there are many studies reporting the existence 
of a (3$\times$4) structure either in coexistence with the 
($\sqrt{3}\times\!\sqrt{3}$) R30$^\circ$ structure or as the only stable 
structure which clearly underlines the importance of chain-chain interactions 
in the film formation mechanism.\cite{guo,tang,voznyy} Though, determination of 
the unit cell structures discussed above were relatively easy through the use 
of diffraction (X-ray, electron, atom) and scanning probe microscopy (mostly 
STM) techniques, determination of the exact arrangement of molecules in these 
unit cells has proved to be a much difficult problem. Hence, a very large 
number of experimental and theoretical studies have been carried out to clarify 
this issue most of which were addressing the questions of where the sulfur atom 
binds on the Au(111) surface and what are the factors that drive the formation 
of c(4$\times$2) 
superlattice.\cite{schreiber1,love,maksymovych,woodruff,vericat2,longo}  To 
this end early computational studies focused on  C1 SAMs to determine sulfur 
binding site since it requires the minimum computational 
power.\cite{maksymovych,vericat2} However, as mentioned above, the 
chain-chain interactions has a significant effect in the film structure and 
computational results obtained for C1 SAMs cannot directly be extended to 
longer chain thiols. Hence in the recent years several density functional 
theory (DFT) studies (some of which also use van der Waals corrections) have 
been employed to study longer chain thiol 
SAMs.\cite{quiroga,carro,nadler,torres,ferrighi,otalvaro,fajin,forster,luque}  
Nevertheless in almost all of these computational studies the longest thiol 
studied was C6 and mostly standing up high density films were examined with 
the aim of determining most favored gold surface reconstruction which seems to 
be RS-Au$_{\textrm{\scriptsize adatom}}$-SR. 

\begin{figure}
\includegraphics[width=8cm]{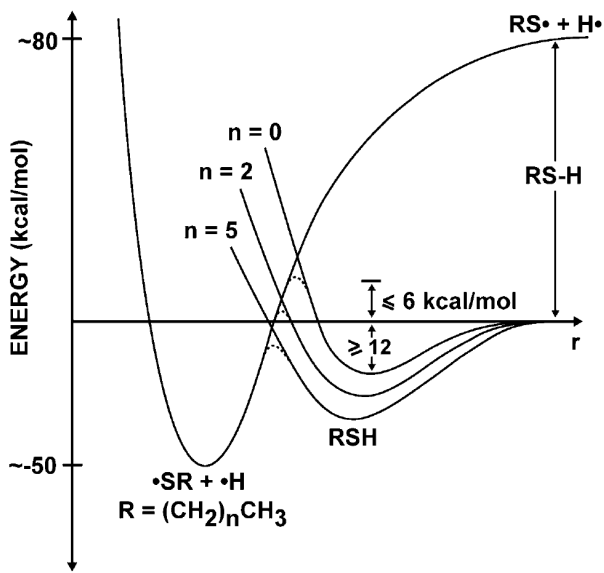}
\caption{Reprinted with permission from Ref~\cite{dubois}.
Copyright 1993, American Institute of Physics.\label{fig1}}
\end{figure}

Lying down (striped) phases, though, are particularly important for film 
formation mechanism, have almost never been studied computationally. Based on 
gas phase studies it is believed that chemisorption of thiols on Au(111) takes 
place through a physisorbed precursor state where the thiol molecules are lying 
down on the surface and relatively 
mobile.\cite{love,dubois,vericat2,schreiber2} With increasing 
chain length not only the physisorption energy increases but also the energy 
barrier between the chemisorbed and physisorbed states 
[(RS-H)$_{\textrm{\scriptsize phys}}$Au $\rightarrow$ RS-Au+1/2H$_2$) is 
suggested to decrease. It was estimated that for chains longer than 6 carbons, 
energy of the transition state lies below the molecular desorption energy (see 
Figure~\ref{fig1}).\cite{dubois} In addition, it was found, based on 
temperature programmed desorption (TPD) measurements, that for chains longer 
than 14 carbons physisorption energy is higher than the chemisorption 
energy.\cite{lavrich} Finally, it should be noted that in all the 
experimental (STM) studies about striped phases  the plausibility of the 
observed unit cell patterns was judged by simply assuming the molecules are 
perfectly parallel to the gold surface in all-trans fashion or have a tilt 
angle different than the standing up phases (see for example references 
[\cite{poirier,qian,toerker,darling,munuera}]).

To address these issues and to help interpret the rich experimental findings on 
striped phases, however, only very few and partial computational studies have 
been performed.\cite{ferrighi,forster,luque} For instance Ferrighi 
and coworkers examined low density C4 SAMs in RS-Au$_{\textrm{\scriptsize 
adatom}}$-SR configuration and found that while with standard DFT (PBE 
functional) the chemisorption energy does not change with the tilt angle of the 
molecules, when dispersion corrections were included (M06-L functional) the 
lying down phase becomes significantly more stable than the standing up 
phase.\cite{ferrighi} Tonigold and Gro\ss~ performed a similar study for thiols 
C1 through C6 and found that with PBE functional chemisorption energy does not 
change significantly (at low density) with changing tilt angle or chain 
length.\cite{forster} However when they used PBD-D3 dispersion correction 
scheme, though there was no significant change in the chemisorption energy of 
standing up molecules with changing chain length, for tilted (lying down) 
molecules it increased significantly with increasing chain length. Finally, 
Luque and coworkers studied C3 SAMs and compared the physisorbed and 
chemisorbed states of standing up and lying down phases at different coverages 
and for different Au surface reconstructions/defects (e.g. adatom, 
vacancy).\cite{luque} Though they considered the van der Waals interactions 
only through the use of a simple universal force field approach, interestingly 
they found for a defect free surface chemisorption energy of a standing up 
molecule to be higher than that of lying down molecule regardless of the 
coverage. On the other hand, for sulfur binding to an Au adatom or on top of an 
Au vacancy site chemisorption strength was strongly depending on the coverage 
and for certain coverage values lying down phases had higher chemisorption 
energy. 

To address the issues summarized above regarding striped phases of thiol SAMs 
here we present a systematic vdW-DFT study of physisorption and chemisorption 
of thiols C1 through C10. First we discuss the physisorption of 
isolated lying down thiols on defect free (unreconstructed) Au(111) surface. 
Then we present the chemisorption energies for the same configurations and 
compare these energies as a function of chain length in order elucidate 
physisorption to chemisorption transition mechanism/energetics. Finally, we 
study the adsorption energies for full monolayer C10 films in lying 
down (11$\times\sqrt{3}$) and C1, C3, C4, C6, C8, and C10 films in standing up 
[($\sqrt{3}\times\!\sqrt{3}$)R30$^\circ$] configurations with the aim of 
elaborating on the experimentally determined striped phase unit cell 
structures.

\section{Theoretical Methods}

\begin{figure}
\includegraphics[width=8cm]{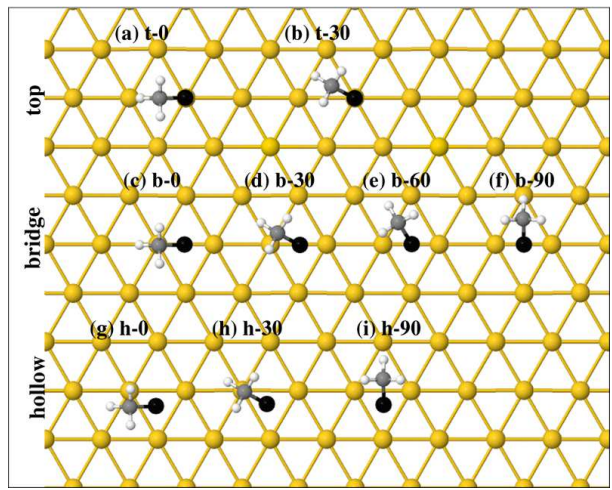}
\caption{Possible initial binding geometry models of an isolated 
alkanethiol on Au(111) surface. For visual convenience methanethiol (C1) is 
depicted as a representative. The labeling starts with the first letter of 
chemisorption site of sulphur atom (in black), then the angle (in degrees) 
between the major axis of the molecule and the gold row follows after 
that.\label{fig2}}
\end{figure}

The minimum energy geometries of alkanethiols on the (111) surface of gold have 
been determined based on density functional theory (DFT) calculations using
VASP.\cite{Kresse1,Kresse2} Electron-ion interactions were included within 
the framework of the projector-augmented wave (PAW)\cite{Blochl,Kresse3} 
method using the plane wave expansion for the single particle states up to a 
cutoff value of 400 eV. 

For weakly interacting organic-organic and metal-organic systems, van der 
Waals forces need to be included in the calculations. Extensive tests have 
been made to get a proper description of the dispersive forces. Meanwhile,
the lattice structure of metals, in which van der Waals interactions are 
negligibly small, should not be distorted. For instance, the experimental 
lattice constant of bulk gold is 4.078 {\AA}.\cite{Wyckhoff} The 
vdW-DF2\cite{Lee_2010} functional gives a value of 4.33 {\AA} while it turns 
out as 4.16 {\AA} with the standard PBE functional. The density dependent 
dispersion correction (dDsC)\cite{Steinmann1,Steinmann2} scheme leads to 
a lattice parameter of 4.11 {\AA} for gold. The dDsC approach not only is 
computationally efficient but also yields reasonably accurate results among the 
other functionals for gold-alkanethiol structures tested here. These findings
are in agreement with a recent benchmarking study of a similar 
platinum-molecular system.\cite{Gautier}  

\begin{table*}[h!]
\caption{The average S-Au bond lengths ($d_{\textrm{\scriptsize S-Au}}$ in 
angstroms), the tilting angles ($\theta$ in degrees) and the heights of the 
carbon atoms at the tip ($h_{tip}$ in angstroms) of isolated alkanethiol 
molecules on Au(111) using both PBE and PBE+dDsC functionals.  
\label{table1}}
\begin{tabular}{l|ccc|ccc||ccc|ccc}
\multirow{2}{*}{} & \multicolumn{6}{c}{Physisorption} & 
\multicolumn{6}{c}{Chemisorption}\\[1mm]
& \multicolumn{3}{c}{PBE} & 
\multicolumn{3}{c}{PBE+dDsC} & 
\multicolumn{3}{c}{PBE} & 
\multicolumn{3}{c}{PBE+dDsC} \\[1mm]\cline{2-13}\\[-3mm]
& $d_{\textrm{\scriptsize S-Au}}$ & $\theta$ & $h_{\textrm{\scriptsize tip}}$ &
$d_{\textrm{\scriptsize S-Au}}$ & $\theta$ & $h_{\textrm{\scriptsize tip}}$ &
$d_{\textrm{\scriptsize S-Au}}$ & $\theta$ & $h_{\textrm{\scriptsize tip}}$ &
$d_{\textrm{\scriptsize S-Au}}$ & $\theta$ & $h_{\textrm{\scriptsize tip}}$ 
\\[1mm]\hline
methane & 2.66 & 25.8 & 4.46 & 2.62 & 22.8 & 4.33 & 
2.46/2.47 & 34.1 & 3.92 & 2.47/2.48 & 33.9 & 3.74 \\
ethane & 2.66 & 21.3 & 4.44 & 2.62 & 20.1 & 4.31 & 
2.47/2.47 & 36.0 & 4.54 & 2.48/2.48 & 33.8 & 4.15 \\
propane & 2.67 & 19.8 & 5.06 & 2.62 & 17.7 & 4.78 &  
2.46/2.47 & 35.4 & 5.41 & 2.48/2.49 & 31.7 & 4.97 \\
butane & 2.67 & 16.7 & 5.10 & 2.62 & 14.5 & 4.70 & 
2.46/2.47 & 32.2 & 5.57 & 2.48/2.49 & 23.0 & 4.84 \\
pentane & 2.66 & 11.8 & 5.08 & 2.63 & 10.4 & 4.78 & 
2.46/2.47 & 30.5 & 6.41 & 2.49/2.49 & 18.7 & 5.04 \\
hexane & 2.67 & 11.9 & 5.19 & 2.63 & 8.4 & 4.66 &  
2.46/2.47 & 24.9 & 6.02 & 2.47/2.48 & 13.6 & 4.77 \\
heptane & 2.70 & 8.8 & 5.15 & 2.66 & 6.8 & 4.63 & 
2.47/2.48 & 16.2 & 5.57 & 2.48/2.49 & 10.5 & 4.65 \\
octane & 2.69 & 6.8 & 4.98 & 2.64 & 5.2 & 4.61 & 
2.48/2.49 & 13.3 & 5.38 & 2.48/2.49 & 9.4 & 4.62 \\
nonane & 2.69 & 5.7 & 5.01 & 2.64 & 4.4 & 4.59 & 
2.48/2.49 & 11.1 & 5.28 & 2.48/2.49 & 8.2 & 4.69 \\
decane & 2.67 & 5.2 & 4.87 & 2.64 & 4.1 & 4.58 & 
2.48/2.48 & 8.7 & 4.98 & 2.48/2.48 & 6.7 & 4.49 \\ 
\end{tabular}
\end{table*}

In order to consider isolated molecules (C1--C10) on Au(111), a number 
of p($n\times m$) slab models with various surface periodicities have been 
constructed, where $n$,$m$ are suitably chosen integers. For instance, a 
single C1 can be thought isolated on a p(4$\times$4) slab. Similarly, a 
p(7$\times$5) cell provides enough room for a C10 molecule to be separated 
from its periodic images. A ($\sqrt{3}\times\!\sqrt{3}$)R30$^\circ$ and a 
(11$\times\!\sqrt{3}$) supercell structure has been formed to simulate C1, 
C3, C4, C6, C8, C10 standing up and C10 lying down full monolayer phases, 
respectively. These supercell models are chosen in accordance with the 
experimentally assigned surface structures.\cite{love,woodruff,qian,guo} All 
computational cells consist of a slab with four layers of 
gold, molecular adsorbates and a vacuum region of at least 12 {\AA} 
thick. Brillouin zone integrations were performed with Methfessel-Paxton 
smearing of 0.1 over appropriate \textbf{k}-point grids which both obey the 
translational symmetry of the corresponding supercell and yield a 10$^{-4}$ eV 
convergence in the total energies. For instance, 6$\times$6$\times$1 and 
3$\times$5$\times$1 $\Gamma$-centered \textbf{k}-meshes were used for 
p(4$\times$4) and p(7$\times$5) real space structures. The geometry 
optimizations were performed self-consistently both with and without the vdW 
corrections by requiring the Hellmann-Feynman forces on each ion in each 
direction to be less than 10$^{-2}$ eV$\cdot${\AA}$^{-1}$. We used the 
corresponding lattice constant of gold obtained for each type of 
exchange-correlation functional in the slab calculations. After achieving a 
full relaxation of the clean surface models, the ionic positions in the bottom 
layer were kept frozen in subsequent molecular adsorption calculations.

The adsorption energy per alkanethiol molecule on Au(111) surface 
can be calculated using,
\[
E_{\rm a}=\left(E_{{\rm C}n+{\rm Au(111)}}-E_{\rm Au(111)}-mE_{{\rm 
C}n}\right)/m\, ,
\]
where $E_{{\rm C}n+{\rm Au(111)}}$ is the total supercell energy of 
Au(111) slab with $m$ number of C$n$ molecules, $E_{\rm Au(111)}$ and $E_{{\rm 
C}n}$ are the energies of the bare Au(111) slab and of a single C$n$ in the gas 
phase, respectively. In chemisorption cases, C$n$ refers to the molecule 
where the hydrogen atom is removed from the thiol group. Adsorption energies, 
$E_{\rm a}$, will be referred to as $E_{\rm p}$ for physisorption and $E_{\rm 
c}$ for chemisorption, in addition ``dDsC" and ``PBE" will be used as 
subscripts to distinguish the adsorption energies calculated by these methods 
when necessary. Finally, we should remark that when adsorption energies are 
compared in the discussion part, always the magnitudes (absolute values) will 
be considered.

\begin{figure*}[h!]
\includegraphics[width=14.5cm]{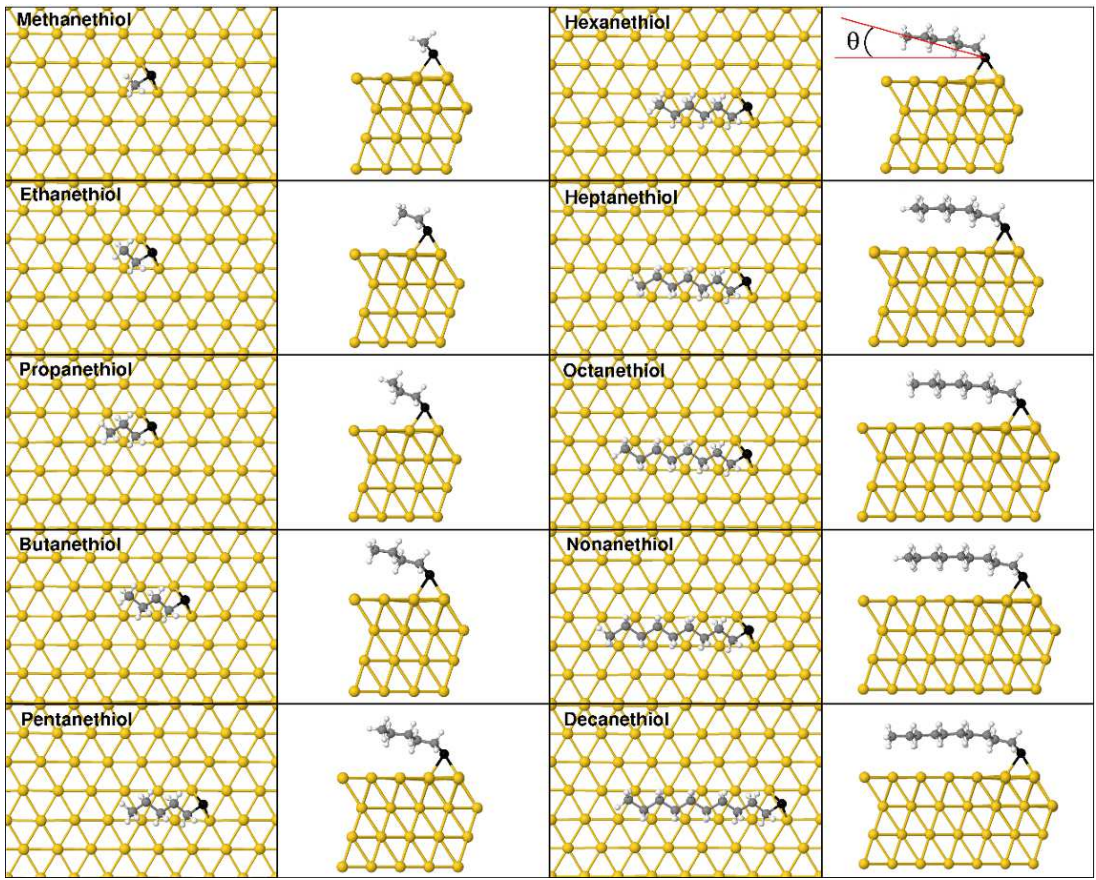}
\caption{Chemisorption geometries of isolated alkanethiols on 
Au(111) surface optimized using the PBE+dDsC vdW corrected DFT calculations. 
The top and side views are presented for C1 up to C10 molecule. The 
tilting angle, $\theta$, is measured between the surface parallel and the line 
connecting the sulphur atom to the last carbon at the tip of the 
corresponding molecule.\label{fig3}}
\end{figure*}

\section{Results and Discussion}

\vspace{5mm}\textbf{Isolated Alkanethiols on Au(111)}\\

Single isolated alkanethiols were considered with all probable initial 
configurations for physisorption and chemisorption on Au(111) surface.
The symmetrically distinct structures are shown in Figure~\ref{fig2} where the 
C1 is chosen as a representative for visual convenience. The 3$p^4$ valance 
property allows sulphur atom to interact with gold more strongly relative to 
the atoms at the carbon chain. Hence, hydrogen deficient thiol drives 
the molecules to the bridge site in all of the geometry optimization 
calculations starting with any of the initial chemisorption configurations 
given in Figure~\ref{fig2}. As a result, the alkanethiols energetically prefer 
the b-90 position with the formation of two S-Au bonds as shown in 
Figure~\ref{fig3}. Inclusion of the vdW corrections has a little effect on the 
binding site and brings slight differences in the structural parameters like 
bond lengths and tilting angles (in Table~\ref{table1}) of isolated molecules 
on gold. Since for longer chains there is significant bending in the molecule, 
the tilting angle is calculated by using the line connecting sulfur atom to the 
last carbon on the chain. In particular, the PBE S-Au bond lengths increase 
from around 2.46 {\AA} for C1 to 2.48 {\AA} for C10 while the PBE+dDsC ones are 
less dependent on the chain size giving a value of 2.48 {\AA} on the average. 
However, the values obtained with these two different functionals converge to 
each other as the chain lengths increase. On the other hand, the tilting of the 
molecules with respect to the gold (111) plane seems to be noticeably smaller 
with the vdW corrections. Moreover, the difference in the tilting angles gets 
bigger for C4-C7 for which long-ranged dispersive forces between the surface 
and the carbon chain becomes important for these midsize molecules. Apparently, 
the height of the tips of the chemisorbed alkanethiols from the surface is 
always larger with the PBE functional. This is not surprising since vdW 
functional is expected to account for dispersion interaction better than PBE 
which in turn results in approaching of the chain tip to the surface, 
especially so for the longer chains that have larger flexibility.

\begin{figure*}[htb]
\includegraphics[width=14.5cm]{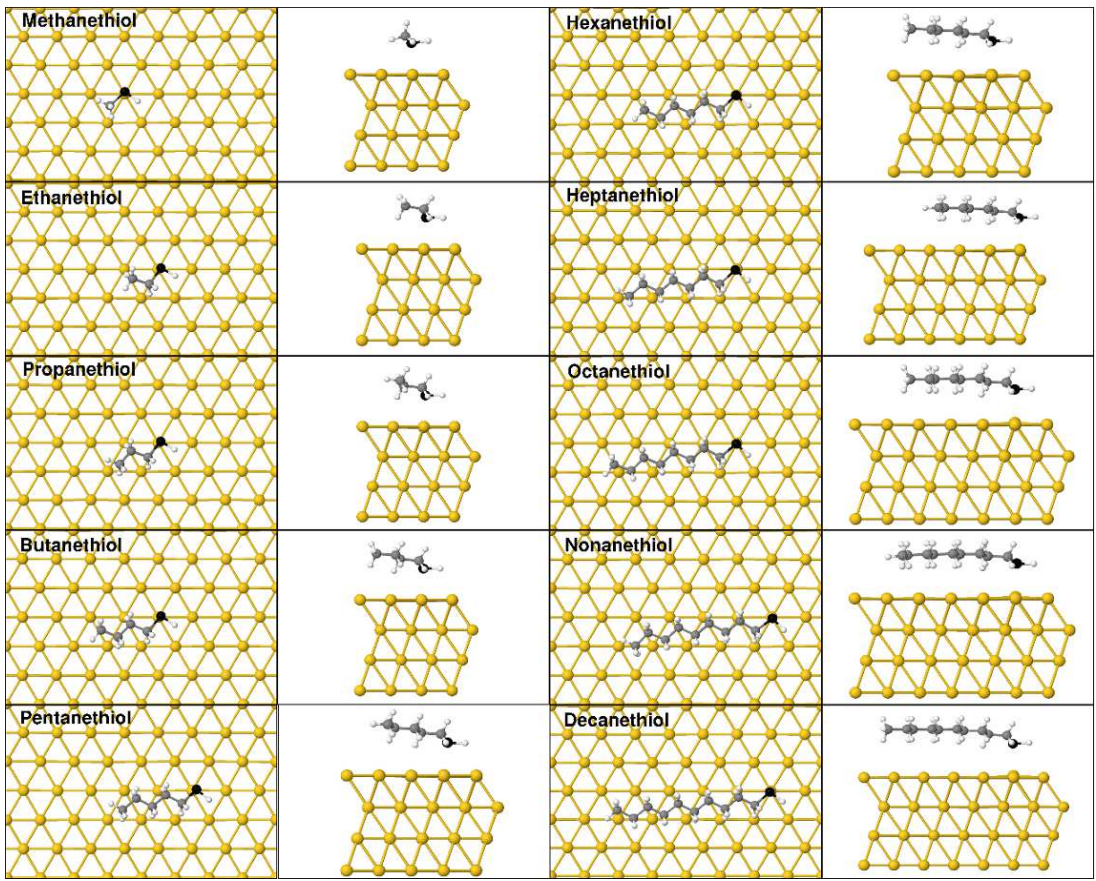}
\caption{Physisorption geometries of isolated alkanethiols on 
Au(111) surface optimized using the PBE+dDsC vdW corrected DFT calculations. 
The top and side views are presented for C1 up to C10 molecule.\label{fig4}}
\end{figure*}

Geometry optimizations were also carried out for the physisorption of 
alkanethiols (in Figure~\ref{fig4}). As a trend, the thiol part energetically 
relax in the top site with the alkyl groups lying in between two adjacent gold 
rows where carbon atoms are attracted to the nearest surface gold atoms. The 
physisorbed molecules always end up with significantly smaller tilting angles 
relative to their chemisorbed counterparts. As expected, the tilting exhibits 
an inverse proportionality with the chain length. In addition, considerable 
bending starting from the S-C bond toward the tip results due to the S-Au 
interaction which especially leads to a bond formation in the chemisorption 
of C7-C10. Molecular bending through the major axis is relatively less in the 
physisorption cases where the S-Au distance is significantly larger. PBE 
functional gives values between 2.66 {\AA} and 2.70 {\AA} for the S-Au distance 
while the vdW corrections yield values between 2.62 {\AA} and 2.66 {\AA}. 

\begin{figure*}
\includegraphics[width=12.5cm]{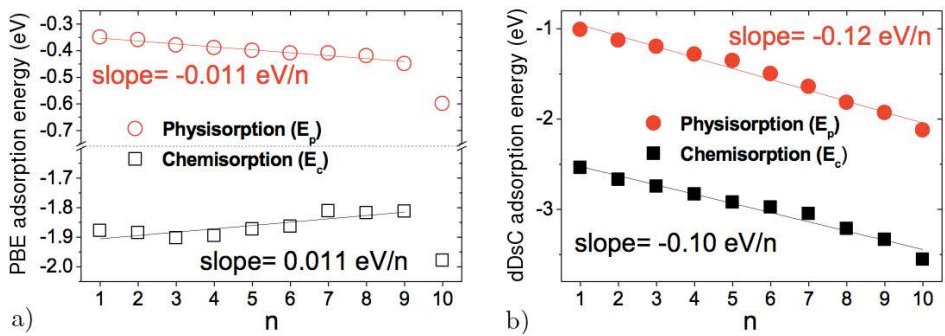}
\caption{Calculated physisorption and chemisorption energy plots of 
isolated alkanethiols on Au(111) with PBE and PBE+dDsC functionals.\label{fig5}}
\end{figure*}

\begin{figure*}
\includegraphics[width=12.5cm]{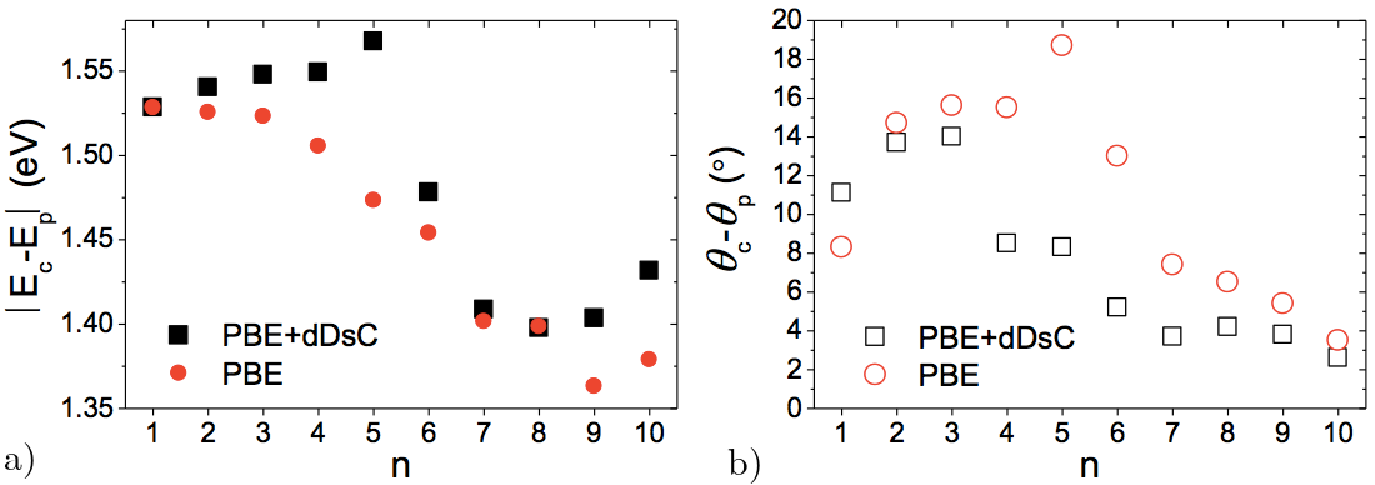}
\caption{Energy and tilting angle differences between chemisorbed and 
physisorbed isolated alkanethiols using PBE and PBE+dDsC 
functionals.\label{fig6}}
\end{figure*}

The calculated physisorption and chemisorption energies are reported in 
Table~\ref{table2}. PBE physisorption energies ($E_{p, 
\textrm{\scriptsize PBE}}$) indicate a weak interaction of the thiol compounds 
with the gold surface. On the other hand, dDsC physisorption energies ($E_{p, 
\textrm{\scriptsize dDsC}}$) are larger and increase significantly with 
increasing chain length due to the long range correlation effects between the 
non-overlapping charge densities. These effects are clearly visible in the 
plots provided in Figure~\ref{fig5} which show the adsorption energies as a 
function of chain length. In case of the PBE functional, physisorption energy 
increases linearly with chain length but with a slope of only -0.011 eV/n, on 
the other hand with the dDsC functional the slope is one order of magnitude 
larger (-0.12 eV/n). One interesting point to note here is that with PBE 
functional C10 physisorption energy does not follow this trend and is higher 
than what the linear fit predicts. When the geometry of physisorbed C10 is 
compared with the other physisorbed thiols (C1-C9) no significant difference 
can be observed which makes it difficult to predict the reason of this deviation 
in the physisorption energy of C10. 

When the chemisorption energies with and without dispersion corrections are 
compared, a trend similar to what is discussed above for physisorption can be 
observed. The chemisorption energy with the PBE functional,
($E_{c,\textrm{\scriptsize PBE}}$), is almost constant (at about -1.9 eV) . If 
a linear fit is forced however, a decrease with increasing chain length can be 
found with a slope of 0.011 eV/n (with a poor R$^2$ value of 0.68). Very 
interestingly C10 chemisorption energy does not follow this trend and is 
significantly higher (in magnitude) than what the linear fit predicts. Since 
this was also the case in the physisorption, the nature of the cause of this 
deviation (in the PBE binding energies of C10) may be the same for both 
physisorption and chemisorption. When the dDsC chemisorption energies are 
examined ($E_{c, \textrm{\scriptsize dDsC}}$), a significant increase (in 
magnitude) can be observed with increasing chain length. This change is pretty 
linear with a slope of -0.10 eV/n. It is not surprising that the slopes of 
chemisorption and physisorption plots with PBE+dDsC (will be referred to as 
$S_c$ and $S_p$ respectively) are similar, since in the chemisorption case the 
Au-S binding energy is more or less independent of the chain length (as 
evidenced by PBE chemisorption energy values) whereas the alkyl chain-gold 
surface interaction increases almost to the same extent it increases in the 
physisorption. 

Experimentally, physisorption of alkane thiols as a function of chain length 
was studied by Scoles group and the desorption energy was found to increase 
with a slope of 0.063 eV/n.\cite{lavrich} This value is about the half of what 
we found here either for chemisorption or physisorption. One reason may be the 
fact that the results we obtained are for isolated molecules whereas the 
experimental values were obtained by desorbing a full monolayer of physisorbed 
film. In general, however, computational studies find the binding energies of 
isolated molecules to be larger (in magnitude) than the binding energies for a 
monolayer.\cite{luque,vargas,cometto,lustemberg} In fact, this is the case in 
this study as well and, as will be discussed in the next section, we found the 
binding energies of a full monolayer of C10 (either physisorbed or chemisorbed) 
to be lower than that of isolated C10 molecules. 

\begin{table}
\caption{Adsorption energies (eV) of isolated alkanethiol molecules 
on Au(111) calculated using both PBE and PBE+dDsC functionals.
\label{table2}}
\begin{tabular}{lcccc}\hline
\multirow{2}{*}{} & \multicolumn{2}{c}{Physisorption} & 
\multicolumn{2}{c}{Chemisorption}\\
& PBE & 
PBE+dDsC & 
PBE & 
PBE+dDsC \\[1mm]\hline
methane & -0.35 & -1.01 & -1.88 & -2.54\\
ethane & -0.36 & -1.13 & -1.89 & -2.67\\
propane & -0.38 & -1.20 & -1.90 & -2.75\\
butane & -0.39 & -1.28 & -1.90 & -2.83\\
pentane & -0.40 & -1.36 & -1.87 & -2.92\\
hexane & -0.41  & -1.50 & -1.86 & -2.98\\
heptane & -0.41 & -1.64 & -1.81 & -3.05\\
octane & -0.42 & -1.82 & -1.82 & -3.22\\
nonane & -0.45 & -1.93 & -1.81 & -3.34\\
decane & -0.60 & -2.13 & -1.98 & -3.56\\ \hline
\end{tabular}
\end{table}

The difference between the chemisorption and physisorption energies as a 
function of chain length is another parameter than can be considered for 
comparing the strength of hydrocarbon chain-gold and sulfur-gold interactions. 
In Figure~\ref{fig6}a  the magnitude of this difference ($E_c$-$E_p$) is 
plotted for both PBE and PBE+dDsC adsorption energies. As can be seen, with 
PBE, the difference decreases (in magnitude) gradually from C1 to C10. However, 
the rate of this decrease is largest in the range of C4 to C7. In case of 
PBE+dDsC adsorption energies, on the other hand, there is no such trend. In 
fact, the difference first increases (in magnitude) slowly up to C5 and then 
suddenly decreases and stays almost constant. This behavior is in fact strongly 
correlated to the tilt angle difference between the physisorbed and chemisorbed 
molecules ($\theta_c$-$\theta_p$) as shown in Figure~\ref{fig6}b. As the length 
of the molecule increases, it becomes more flexible and the tilt angle of the 
chemisorbed state (which is constrained for short chains by the Au-S bond) 
approaches to that of the physisorbed state. This transition takes place at 
chain length of 5-6 carbon atoms as clearly visible in the plots.

\vspace{5mm}\textbf{Alkanethiol Monolayers on Au(111)}\\

\begin{figure*}[htb]
\includegraphics[width=14cm]{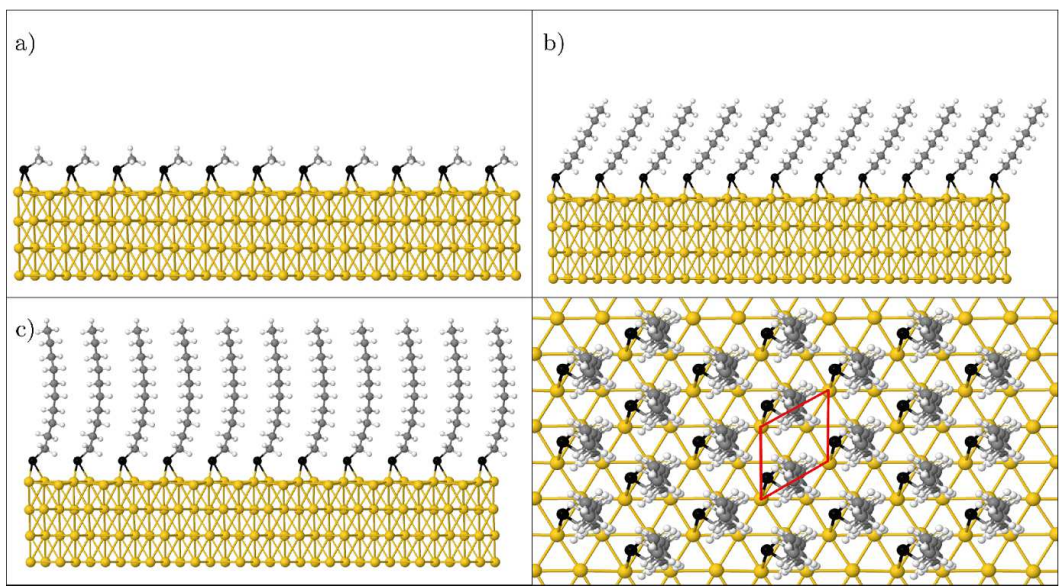}
\caption{One full monolayer of a) C1, b) C6, and c) C10 on 
Au(111) with ($\sqrt{3}\times\!\sqrt{3}$)R30$^\circ$ optimized with 
PBE+dDsC vdW corrected DFT functional. Surface unit cell is 
indicated in the bottom right panel (top view of C10 SAM). \label{fig7}}
\end{figure*}

In order to compare the binding characteristics of the alkanethiols in full 
monolayers with their isolated phases on gold, we modeled the experimentally 
observed ($\sqrt{3}\times\!\sqrt{3}$)R30$^\circ$ surface cell for C1, C3, C4, 
C6, C8 and C10. As discussed in the introduction section, for full monolayers of 
standing up thiols, the existence of ($\sqrt{3}\times\!\sqrt{3}$)R30$^\circ$ 
structure has been shown by many experimental studies. There are many 
disscussions regarding the nature of this unit cell, such as adsorption site of 
the sulphur atoms, gold surface reconstructions and chain-chain interactions. 
Here our aim is not to address all of these issues. Instead we want to focus on 
the chain-chain interactions on an unreconstructed surface and discuss/compare 
the chain length dependence of adsorption energies for isolated and full 
monolayer thiols. Probable initial configurations were taken into account 
including top, bridge, and hollow sites and optimized using the vdW corrected 
PBE+dDsC functional (in Figure~\ref{fig7}). As the chain length increases tilt 
angle of the molecules increases gradually from 29$^\circ$ for C1 up to 
80$^\circ$ for C10 SAMs. More interestingly, the thiolates start to bend as the 
carbon chain length increases. In particular, C10 exhibits a significant 
molecular bending at its ($\sqrt{3}\times\!\sqrt{3}$)R30$^\circ$ SAM structure 
as shown in Figure~\ref{fig7}c. The binding sites and the Au-S bonds are not 
affected from the molecular size as given in Table~\ref{table3}. On the other 
hand, the S-C bond angle with respect to the surface plane increases from 
29.1$^\circ$ for C1 to 37.2$^\circ$ for C3 and stays approximately constant up 
to C10. Moreover, longer molecules are bent more between their 
thiol group and alkyl chains. This can be explained by the existence of two 
different interaction strengths. The thiol-gold interaction is much stronger in 
comparison with the long-ranged intermolecular interactions which are 
essentially between the alkyl parts of neighboring molecules. The anchoring 
structure of the thiol part rather seems to be less affected by the carbon 
chain length. As a result, weaker intermolecular interactions signify the 
important role of the alkyl size on the final geometries. From a theoretical 
point of view, that necessitates the use of vdW corrected functionals in the 
DFT calculations (Table~\ref{table4}).

\begin{figure*}
\includegraphics[width=12.5cm]{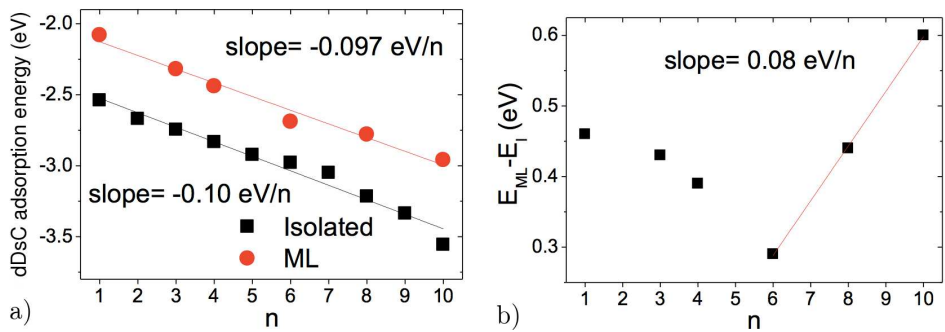}
\caption{Chemisorption energy plots of isolated and full 
monolayer [($\sqrt{3}\times\!\sqrt{3}$)R30$^\circ$] alkanethiols  
on Au(111) optimized with vdW corrected DFT
calculations.\label{fig8}}
\end{figure*}

The PBE+dDsC calculations show that the binding energy per molecule in a full 
monolayer ($E_{\textrm{\scriptsize ML}}$) increases with the increasing length 
of the carbon chain as plotted in Figure~\ref{fig8}a comparatively with the 
binding energies of isolated molecules ($E_{\textrm{\scriptsize I}}$). In fact, 
the chain length dependence of $E_{\textrm{\scriptsize ML}}$ is pretty linear 
and have a slope of -0.097 eV/n (with R$^2$ of 0.97). This is expected since, 
as the molecule length increases the vdW interactions between the chains get 
stronger. Interestingly, however, the slope for standing up monolayer, is 
almost equal to the slope for lying down isolated molecules. Increasing binding 
strength with chain length was also observed by Salvarezza group who studied 
monolayers of C1, C4 and C6 by using optB88-vdW functional to account for the 
vdW interactions.\cite{carro}  However, another probable contribution to this 
increase in the binding energies can be that, for longer chain thiolates, there 
are relatively more unoccupied molecular energy levels available near the Fermi 
energy of the metal surface to allow a larger amount of charge transfer from 
the 11 $d$-states of gold to the empty frontier molecular levels. 

\begin{table}
\caption{Bond lengths ({\AA}), tilting angles (in degrees) and average binding 
energies per molecule (eV) of various alkanethiols at full monolayer coverage 
on gold, calculated using the PBE+dDsC method.
\label{table3}}
\begin{tabular}{lccccc}
\multicolumn{6}{c}{Au(111)-($\sqrt{3}\times\!\sqrt{3}$)R30$^\circ$} 
\\[1mm]\hline
& $d_{\textrm{\scriptsize S-Au}}$ & $d_{\textrm{\scriptsize S-C}}$ & 
$\theta_{\textrm{\scriptsize S-C}}$ & $\theta$ & E$_a$ \\[1mm]\hline
methane & 2.50/2.51 & 1.82 & 29.1 & 29.1 & -2.08 \\
propane & 2.49/2.49 & 1.83 & 37.2 & 49.8 & -2.32 \\
butane & 2.48/2.49 & 1.82 & 37.6 & 52.9 & -2.44 \\
hexane & 2.48/2.49 & 1.82 & 37.6 & 56.3 & -2.69 \\
octane & 2.47/2.50 & 1.83 & 37.8 & 69.4 & -2.78 \\
decane & 2.48/2.50 & 1.83 & 37.7 & 82.5 & -2.96 \\ \hline
\end{tabular}
\end{table}

\begin{table}
\caption{Average adsorption energies per decanethiol (eV) 
in the striped phase at full monolayer coverage on gold, calculated 
using the PBE+dDsC method.
\label{table4}}
\begin{tabular}{cccc}
\multicolumn{4}{c}{Decanethiol/Au(111)-(11$\times\sqrt{3}$)} \\[1mm]
\hline \\[-4mm] 
striped phase& E$_a$ & $d_{\textrm{\scriptsize S-Au}}$ & 
$h_{\textrm{\scriptsize 
tip}}$ 
\\[1mm]
\hline \\[-3mm]
physisorption & -1.32 & 3.89/2.68 & 4.56/4.56 \\[1mm]
chemisorption & -3.03 & 2.43/2.57 & 4.56/4.56\\[1mm] \hline
\end{tabular}
\end{table}

\begin{figure}[htb]
\includegraphics[width=7cm]{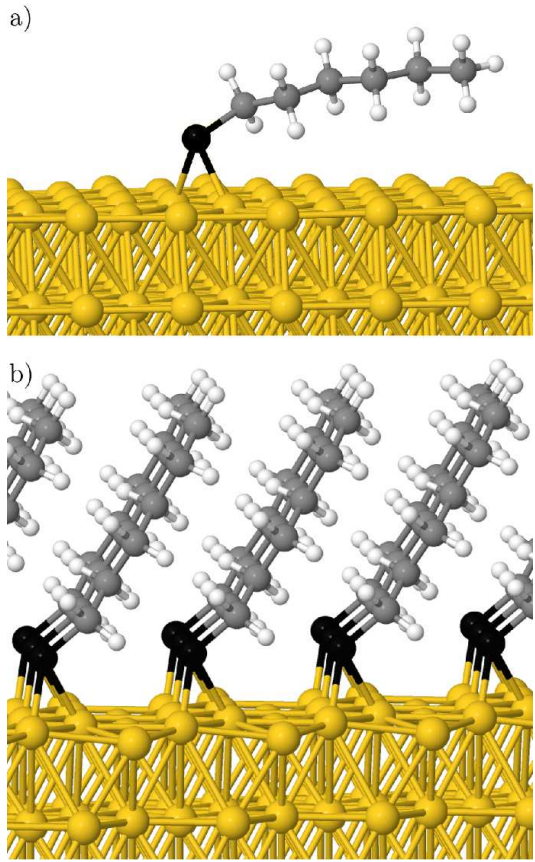}
\caption{Comparison of surface structures of a) isolated and b) full 
monolayer hexanethiol on Au(111) optimized with vdW corrected DFT
calculations.\label{fig9}}
\end{figure}

To have a better interpretation of this chain length dependence of 
$E_{\textrm{\scriptsize ML}}$, 
binding energies of the isolated molecules, $E_{\textrm{\scriptsize I}}$, can 
be compared with $E_{\textrm{\scriptsize ML}}$. As can be seen in 
Figure~\ref{fig8}b, $E_{\textrm{\scriptsize I}}$ is always larger (in 
magnitude) than the corresponding $E_{\textrm{\scriptsize ML}}$, however their 
difference ($E_{\textrm{\scriptsize ML}}$-$E_{\textrm{\scriptsize I}}$) does 
not have a linear dependence on the chain length. Decreasing of the adsorption 
energy of thiols with increasing coverage was also observed in several other 
computational studies~\cite{luque,vargas,cometto,lustemberg} and this is 
generally attributed to the gold surface reconstructions/relaxations becoming 
more difficult at higher coverages. Our results indicate a similar mechanism: 
Each sulphur atom at the bridge site pulls two adjacent gold atoms by $\sim$0.3 
{\AA} up from the (111) plane. Especially, at the full monolayer coverage, 
closely packed alkanethiols put an additional stress on the gold surface. As a 
result, the remaining gold atoms on the upper plane, which are not coordinated 
with any sulphur, are forced to sink a little down leading to a corrugated 
surface look (e.g. as recognized from Figure~\ref{fig9} for C6 in its isolated 
and SAM phases). Therefore, surface relaxation becomes more difficult relative 
to isolated cases. With these in mind, the chain length dependence of 
($E_{\textrm{\scriptsize ML}}$-$E_{\textrm{\scriptsize I}}$) can be explained 
as follows: In the isolated case, for C1 to C6, the molecules have high tilt 
angle and hence low alkyl chain-gold surface interaction. As a result, in the 
ML case these molecules to do not lose much stabilization due to the absence of 
alkyl chain-gold surface interaction (in the ML). In fact, the stabilization 
due to chain-chain interactions present in the ML case can more than compensate 
the lost alkyl chain-gold surface interaction (present in the isolated case). 
However, this (chain-chain interactions) is still not strong enough to 
compensate for the destabilization due to gold surface stress in the ML. Hence,
($E_{\textrm{\scriptsize ML}}$-$E_{\textrm{\scriptsize I}}$) though always 
positive, decreases from C1 to C6, since increasing chain length means stronger 
chain-chain interactions in the ML.  For alkane chains longer than 6 carbons, 
however, in the isolated case there is significant alkyl chain-gold surface 
interaction strength and the chain-chain interactions in the ML is not strong 
enough to compensate the lost alkyl chain-gold surface interactions. Hence, 
($E_{\textrm{\scriptsize ML}}$-$E_{\textrm{\scriptsize I}}$) starts to increase 
after C6 perfectly linearly with a slope of 0.08 eV/n and gets its largest 
value for C10.

\begin{figure*}[htb]
\includegraphics[width=14.5cm]{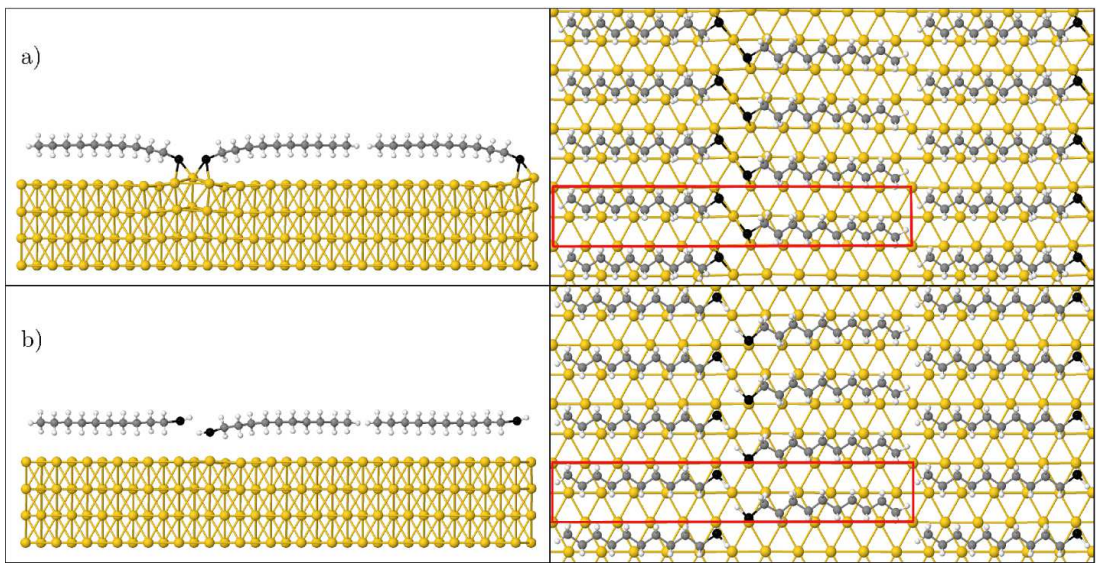}
\caption{One full monolayer of decanethiol (C10) on Au(111) with 
($11\times\!\sqrt{3}$) structure optimized using the PBE+dDsC vdW corrected DFT 
functional. Surface unit cell is indicated on the right panel (top 
view).\label{fig10}}
\end{figure*}

To be able to compare physisorption and chemisorption also for full monolayer 
coverage films and to provide theoretical insight to the striped phase unit 
cell models reported in the literature we investigated the  
Au(111)-(11$\times\!\sqrt{3}$) structure of decanethiol as well by PBE+dDsC 
calculations. The optimized geometries for both physisorption and chemisorption 
are shown in Figure~\ref{fig10}. In the chemisorption case, the sulphur atoms 
lift the surface gold atom up by 0.65 {\AA}, whereas in the physisorption 
case, one of the sulphur atoms approaches the surface significantly, in order 
for the molecules to fit in the unit cell. Other than these differences the 
arrangement of the carbon chain in the unit cell is identical in both cases. 
Our results indicate that (11$\times\!\sqrt{3}$) unit cell models reported in 
the literature (with flat lying down C10 molecules) were in fact pretty 
accurate even though they were not based on any computational study. One 
interesting point to note here is that even though we have not considered 
reconstructed gold surfaces in here (like gold adatoms as discussed in the 
introduction), the interaction of the long C10 chain with the gold surface is 
strong enough to force the molecule to an almost perfectly flat configuration 
by lifting one gold atom from the surface in the chemisorption case. Hence 
(though not an adatom) a reconstruction similar to RS-Au$_{\textrm{\scriptsize 
adatom}}$-SR model reported in the literature~\cite{guo,mazzarello} takes place 
on the surface. 

When the energetics of structures is considered, it can be seen that 
chemisorbed monolayer is much more stable than the physisorbed one, as 
expected. As in the case of ($\sqrt{3}\times\!\sqrt{3}$) monolayer, the 
chemisorption energy of (11$\times\!\sqrt{3}$) structure 
($E_{c,(11\times\sqrt{3})}$=-3.03 eV) is lower 
(in magnitude) than that of the isolated C10 ($E_{c,\textrm{\tiny C10}}$=-3.56 
eV). This decrease is most probably due to the energy spent on lifting one gold 
atom (per unit cell) from the surface. Interestingly, however, the binding 
energy of the physisorbed (11$\times\!\sqrt{3}$) monolayer 
($E_{p,(11\times\sqrt{3})}$=-1.32 eV) is also lower (in magnitude) than that of 
the physisorbed isolated C10 (-3.03 eV). The difference in this latter case is 
much larger and is probably due to the short S-Au distance of one of the 
molecules in the unit cell, which is energetically not favored. Nevertheless, 
$E_{p,(11\times\sqrt{3})}$ is pretty close to the experimentally predicted 
value of 1.1 eV by Scoles group.\cite{lavrich} When the chemisorption energies 
of C10 in the standing up ($\sqrt{3}\times\!\sqrt{3}$) and in the lying down 
(11$\times\!\sqrt{3}$) structures are compared, the latter comes out as being 
slightly more stable in agreement with the previous theoretical 
studies.\cite{ferrighi} This result indicates that alkyl chain-gold surface 
interactions are stronger than the chain-chain interactions supporting the 
discussion made above regarding chain length dependence of 
($\sqrt{3}\times\!\sqrt{3}$) adsorption energies. 

For extrapolating the $S_c$ and $S_p$ values (discussed above for lying down 
isolated molecules) to monolayer coverage in order to have a better comparison 
with the experimentally reported value, we used the ratio of 
(11$\times\!\sqrt{3}$) chemisorption energy to that of isolated lying down C10 
(calculated with PBE+dDsC). When $S_c$ and $S_p$ are divided by this ratio 
(1.17) the resulting corrected slopes, -0.10 eV/n and -0.085 eV/n, are, 
however, still larger (in magnitude) than the experimental value (0.063 eV/n). 
This inconsistency with the experimental slopes and the apparent difference 
between the chemisorption energies reported here for 
($\sqrt{3}\times\!\sqrt{3}$) phases and the experimental values reported in the 
literature (determined by temperature programmed desorption studies, TPD) may 
have common reasons and deserve further elaboration. Experimentally, desorption 
energies of full monolayer chemisorbed alkanethiols are determined to be about 
1.30 eV and do not change with chain 
length.\cite{lavrich,albayrak,albayrak2,albayrak3} Though 
some higher energy desorption features for long alkanethiols (as high as 1.7 eV 
for C16) were also reported, these energies were not chain length 
dependent.\cite{lavrich,albayrak,albayrak2,albayrak3,nuzzo}  However, our 
results indicate much larger energies and more importantly an increase with the 
chain length. To resolve this ``inconsistency" it should be kept in mind that 
during TPD measurements the temperature of the film is ramped and desorption 
takes place gradually. Hence, before desorbing from the surface (before Au-S 
bond is broken), molecules may detach themselves from the island borders and 
diffuse on the surface or change their configuration (since the gold potential 
energy surface is pretty smooth as evidenced by many previous computational 
studies). Such a detachment from island edges than could eliminate the chain 
length dependence of the desorption energies for the chemisorbed phases. 

\section{Conclusions}
The structures and energetics of isolated and monolayer phases of alkanethiols 
on gold (111) surface have been investigated by density functional theory 
calculations. In order to put forward the role of dispersive forces on these 
metal-organic systems, vdW corrections have also been included in the 
calculations using the density dependent dDsC scheme. In all cases, considered 
in this study, physisorption of alkanethiols (C1-C10) ends up with the S atom 
being on top of a surface Au atom. However, chemisorption leads to strong 
binding at the bridge site with the formation of two almost equivalent Au-S 
bonds. While, PBE results indicate negligible dependence of the binding 
energies on chain length, dDsC results show an increase in the binding strength 
with increasing chain length with almost the same rate for both chemisorption 
(-0.10 eV/n) and physisorption (-0.12 eV/n). The difference between the 
chemisorption and physisorption energies and tilt angles indicate that after a 
chain length of 6 carbons the molecules become flexible enough to interact with 
the gold surface even in the chemisorbed configuration (where the molecular 
geometry is constrained by the S-Au bond). It is interesting to note that this 
chain length (n=6) coincides with the estimated chain length for which the 
energy of the transition state between chemisorbed and physisorbed states lies 
below the molecular desorption energy (see Figure~\ref{fig1}).

In case of standing up full monolayers the chemisorption energy increases with 
increasing chain length due to chain-chain interactions with a rate of about 
(-0.1 eV/n). However, regardless of the chain length chemisorption energy of 
the full monolayers are always lower (in magnitude) than that of the isolated 
(lying down) molecules, due to the gold surface stress present in the full 
monolayer films. The difference between full monolayer and isolated 
chemisorption energies, however, starts increasing with a rate of 0.08 eV/n 
after n = 6, indicating that the alkyl chain-gold surface interaction is 
stronger than chain-chain interactions for alkanethiols with n $>$ 6. The fact 
that chemisorption energy for the striped C10 monolayer is slightly larger (in 
magnitude) than that of the standing up C10 monolayer is also in agreement with 
this conclusion. In its striped phase C10 lies perfectly parallel to the 
surface lifting a gold atom up, in agreement with the proposed models for this 
film in the literature.

\bibliography{Alkanethiol_SAMS}
\bibliographystyle{apsrev4-1}

\end{document}